\documentclass[a4paper,11pt]{article}
\pdfoutput=1 

\usepackage{jcappub} 

\usepackage[T1]{fontenc} 
\usepackage{makecell}
\usepackage{multirow}

\title{\boldmath The determination capability of potential neutrinos from gravitational wave sources and contributions of extra detector at the future reactor neutrino experiment}

\author[a,b]{Zhaokan Cheng,}
\author[a]{Jingbo Zhang,}
\author[b]{Chan Fai Wong}
\author[b]{and Wei Wang}

\affiliation[a]{School of Physics, Harbin Institute of Technology,\\92 West Dazhi Street, Nan Gang District, Harbin, 150001, P. R. China}
\affiliation[b]{School of Physics, Sun Yat-Sen University,\\No. 135, Xingang Xi Road, GuangZhou, 510275, P. R. China}

\emailAdd{14b911018@hit.edu.cn}
\emailAdd{jinux@hit.edu.cn}
\emailAdd{wongchf@sysu.edu.cn}
\emailAdd{wangw223@sysu.edu.cn}

\abstract{After several gravitational wave transients were discovered since 2015, studying neutrino signals coincident with the gravitational wave events now becomes an important mission for the existing neutrino experiments. Unfortunately, no candidate neutrinos have been found yet.
This article introduces a method to find the neutrino excess at the future reactor neutrino experiment (such as JUNO and RENO-50), which can be adopted to search for the potential neutrinos from gravitational wave sources.
According to our calculations and simulations, the non-detection of $\bar\nu_e$ associated with gravitational waves at the nominal JUNO experiment gives rise to the $\bar\nu_e$ signal sensitivity at 90$\%$ confidence level (C.L.), $\mu_{90}$ = 2.44. 
This corresponds to the range of neutrino fluence on the Earth around 6 $\times$ 10$^{10}$ cm$^{-2}$ to 4 $\times$ 10$^{10}$ cm$^{-2}$ with neutrino energy range from 1.8 MeV to 120 MeV at monochromatic energy spectrum assumption.
Based on certain popular models which describe the gravitational wave sources, we calculate the corresponding fluence ($F_{UL}^{90}$), which is around 1 - 3 $\times$ 10$^{8}$ cm$^{-2}$ for both monochromatic energy spectrum assumption and Fermi-Dirac energy spectrum assumption.
Then we convert $F_{UL}^{90}$ into the detectable distance ($D_\text{UL}^{90}$), about 1 - 3 Mpc for two assumptions, with the predicted luminosities in these known models. Compared with the KamLAND experiment, the sensitivity of the (future) JUNO-like experiment is expected to improve by a factor of twenty. 
To further improve the sensitivity, we discuss the potential benefits from an extra detector, with different target masses and baselines. We investigate how can an extra detector improve the sensitivity. Particularly, there will be around 38\% sensitivity improvement and around 28$\%$ detectable distance increasing if the extra detector is designed to be identical to the original JUNO detector with the same baseline (53km). 
On the other hand, instead of building an extra detector, if we combine the JUNO experiment with the RENO-50 experiment, the sensitivity will also be significantly improved.
}

\begin{document}
\maketitle
\flushbottom

\section{Introduction}\label{sec1}

In 1987, a neutrino burst in total of 24 events was first discovered by Kamiokande II~\cite{Hirata:1987hu}, IMB~\cite{Bionta:1987qt} and Baksan experiment~\cite{Alekseev:1987ej}, which was confirmed from a supernova (SN1987A) explosion.
After this unprecedented observation, neutrino astronomy has been pushed into a new era and astrophysical neutrino has been developed into an observable stage (situation).
Afterwards, Super-Kamiokande experiment reported neutrinos arised from interaction between cosmic rays and atomsphere~\cite{Fukuda:1998mi}, while SNO experiment obtained solar neutrino flux precisely~\cite{Ahmad:2001an}. 
All these observations promoted the detection of extraterrestrial neutrinos. 
 
In the past two decades, a series of outstanding neutrino detectors and telescopes have been built and the astrophysical neutrinos could make a breakthrough in astrophysics. Particularly, after the first gravitational wave event (GW150914) has been detected~\cite{Abbott:2016blz}, the multi-messenger observations including neutrinos have been consistently gestated and developed.
As neutrino only involves the weak interaction, it could be a good candidate to uncover the dynamics of astronomical phenomena. 
Associating with the other astrophysical messengers, a search for conincident neutrino begins to be performed at neutrino experiments, successively.

Up to now, 11 gravitational wave (GW) transients had been confirmed one-by-one in the Laser Interferometer Gravitational-Wave Observatory (LIGO)~\cite{LIGOScientific:2018mvr}, while 2 gamma-ray bursts, with a delay time of $\sim$ 0.4 s after GW150914 and $\sim$ 1.7 s after GW170817, had been independently detected by the $Fermi$ Gamma-Ray Burst Monitor (GBM)~\cite{Connaughton:2016umz,GBM:2017lvd}, too.
Subsequently, neutrinos associated with these astrophysical signals are considered as important targets in the multi-messenger analyses. Several neutrino experiments and telescopes, such as ANTARRES and IceCube~\cite{Adrian-Martinez:2016xgn,ANTARES:2017iky,ANTARES:2017bia}, KamLAND~\cite{Gando:2016zhq}, Super-Kamiokande~\cite{Abe:2016jwn,Abe:2018mic} and Borexino~\cite{Agostini:2017pfa}, successively performed a series of searches for these coincident neutrinos at different flavors and in different energy ranges. However, no neutrino candidates were found up to now.  
Nevertheless, the non-detections of these experiments can always provide constraints to the neutrino fluence and the total integrated luminosity of astrophysical sources, which are important in astronomy.

Among the above experiments, KamLAND, as a reactor neutrino experiment, is mainly designed to search for $\sim$ MeV neutrinos and antineutrinos~\cite{Suzuki:2014woa}. Within a background rate of several events per day, KamLAND is significantly sensitive to the low-energy ($\sim$ MeV) neutrinos associated with gravitational wave events. 
However, the target mass of the detector is just around 1 kton and it can hardly provide enough statistics to measure the neutrino signals from the gravitational wave sources at the level of $\sim$ Mpc distance.  
On the other hand, the future medium baseline reactor neutrino experiments, JUNO~\cite{An:2015jdp} and RENO-50~\cite{Kim:2014rfa} are proposed to build detector(s) with around 20 ktons target mass. Such large detectors are expected to significantly improve the detection capabilities of the low-energy neutrinos, especially from remote astrophysical sources.  
They are considered as two new eyes to monitoring the outer space and will make contributions to the multi-messenger observations in the future. 

In this article, we refer to the nominal JUNO experiment and perform a series of numerical simulation to estimate backgrouds at the JUNO detector. Using these simulated backgrouds, we investigate the temporal correlation between two adjacent events and analyze the detection sensitivity of the low-energy electron antineutrinos from gravitational wave sources.   
Up to now, LIGO and Virgo have measured two kinds of gravitational wave signals. One is emitted from the binary black holes (BBH) mergers and the other is emitted from the binary neutron star (BNS) mergers~\cite{LIGOScientific:2018mvr}. In addition, the black hole and the neutron star (BH-NS) mergers are expected to be able to produce gravitational waves as well~\cite{Taniguchi:2005fr,Lee:1999kcb}.
Regarding to models referred to BNS~\cite{Sekiguchi:2015dma,Foucart:2015gaa,Sekiguchi:2016bjd}, BH-NS~\cite{Deaton:2013sla,Foucart:2015vpa} and BBH~\cite{Caballero:2015cpa} sources, we discuss the constraints on the $\bar\nu_e$ fluence and the detectable distance of sources at the JUNO experiment.

This paper is organized as follows. In section~\ref{sec2}, we briefly introduce the medium baseline ($\sim$ 50 km) reactor neutrino experiments in the future, JUNO~\cite{An:2015jdp} and RENO-50~\cite{Kim:2014rfa}. In section~\ref{sec3}, we perform a Monte Carlo (MC) event simulation based on the backgroud rate at the nominal JUNO experiment and discuss the time-correlation between adjacent events to monitor neutrino excess. 
In section~\ref{sec4}, we analyze the detection capability of the nominal JUNO-like detector and estimate the detection sensitivity of neutrinos from potential gravitational wave sources without any detected signals. Within two assumed energy spectra, we then calculate the constraints on the fluence and the detectable distance for gravitational wave sources at different models. 
In section~\ref{sec5}, scanning the contributions of detection capability provided by one extra detector, we discuss the status that one additional JUNO-like detector is taken in the nominal JUNO experiment and evaluate the combined results of the sensitivity with the nominal JUNO and RENO-50 experiment together. Finially, we make a summary in section~\ref{sec6}.

\section{The medium baseline reactor neutrino experiment}\label{sec2}
\subsection{The Experimental Setup}
The medium baseline reactor neutrino experiments, such as JUNO~\cite{An:2015jdp} and RENO-50~\cite{Kim:2014rfa}, are designed to determine the neutrino mass hierarchy via precise spectral measurement of the electron antineutrino ($\bar\nu_e$) oscillations. 
JUNO project was approved by Chinese Academy of Sciences in February 2013. It is expected to start data taking in 2020. On the other hand, RENO-50 has obtained R\&D funding and plans to start the construction of facility and detector in 2016 - 2021. It aims to start data taking in 2022.
The main goals of both JUNO and RENO-50 are to identify the neutrino mass hierarchy with $\sim$ 3 $\sigma$ C.L.. The idea is that a large liquid-scintillator detector ($\sim$ 10 - 20 ktons) with excellent energy resolution (3\%/$\sqrt{E}$), locating at around 50 km from the reactor core(s), could observe the subdominant oscillation patterns and thus extract the MH information from the spectral distortion. 

To resolve the neutrino mass hierarchy in reactor neutrino experiment(s), besides extraordinary energy resolution, large statistics is also required. Hence the detectors of this kind of experiments are extremely huge (around 20 times larger than KamLAND) and the data-taking are expected to last for at least 6 years. With such large detector and so long data-taking, the medium baseline reactor neutrino experiment(s) is likely to provide an ideal platform for the study of astrophysical neutrinos.
For example, regarding to the supernova burst neutrinos, a JUNO-like detector is expected to register about 5000 inverse beta decay events for a typical galactic distance of 10 kpc and typical supernova parameters~\cite{An:2015jdp}. Therefore, JUNO and RENO-50 are expected to be comparable to Super-Kamiokande. The combined analyses from these detectors, together with the measurements from gravitational wave detectors, are expected to be able to provide a detailed picture on these astrophysical multi-messengers.

The JUNO detector consists of a central detector, a water Cherenkov detector and a muon tracker. The central detector is a liquid scintillator (LS) detector within a target mass of 20 kton at a $\sim$ 34.4 m fiducial volume. It is built in the Jinji town located at $\sim$ 52.5 km from the Yangjiang Nuclear Power Plant (NPP) and Taishan NPP, where 10 rector cores offer the thermal power of $\sim$ 35.8 GW$_\text{th}$ in total. 
In our calculation, we use the nominal JUNO experimental configurations except that we simplify the study by assuming an ideal single powerful reactor core which is equivalent to the 10 reactor cores in the real case.
Additionally, the RENO-50 is still under proposal but the design of the detector is expected to be similar with JUNO~\cite{Kim:2014rfa}. 

\subsection{Backgrounds}
In order to search for neutrino signals from potential gravitational wave sources in reactor neutrino experiments, we have to study the backgrounds very carefully. Generally, backgrounds are dependent on the location and type of the detector, as well as the signal channel used. Furthermore, backgrouds are determined by the energy range and the selection criteria adopted in the analysis.
As the typical LS detector, JUNO and RENO-50 detector are primarily designed to detect electron antineutrinos via the the inverse beta-decay (IBD) reaction: $\bar\nu_e + p \rightarrow e^+ + n$. Therefore, we concentrate on the analysis on $\bar\nu_e$ signals. To perform the analysis associated within gravitational waves, we make the choice on the energy range from 1.8 MeV to 120 MeV, similar with the KamLAND~\cite{Gando:2016zhq}.
Therefore, the major backgroud is the electron antineutrinos ($<$ 12 MeV) from the reactor core(s) in our analysis. In addition, backgrounds at the reactor neutrino oscillation analysis~\cite{Adey:2018zwh}, such as accidental backgroud, $^8$He/$^9$Li, fast neutrino (FN) and ($\alpha$, $n$) backgroud are the important backgrounds as well.
The predominant background at above 12 MeV (exceeding reactor neutrino energy) energy is the fast neutron induced by cosmic rays. Besides, the geoneutrino background has a significant event rate so that it cannot be ignored. 
However, the events of solor neutrino and the atomspheric neutrino background are negligible at this energy range~\cite{An:2015jdp}.

For the nominal JUNO detector, the reactor neutrino backgrouds are emitted from the Yangjiang NPP and Taishan NPP, with effective baselines of $\sim$ 52.5 km and a total thermal power of 35.8 GW$_\text{th}$, Daya Bay NPP with a baseline of $\sim$ 215 km and a total thermal power of 17.4 GW$_\text{th}$, and Huizhou NPP with a baseline of $\sim$ 265 km and a total thermal power of 17.4 GW$_\text{th}$, respectively.
Like the reactor neutrino, geoneutrino is one kind of neutrino produced from radioactive decays of Th and U inside the Earth that is treated as background in our analysis.

Besides the reactor neutrino and geoneutrino, the remanent backgrounds are basically originated from three other sources, natural radioisotopes, cosmogenic isotopes and cosmic muon. 
The accidental backgroud is mainly resulted from tree type of random coincidence: (radioactivity, radioactivity), (radioactivity, cosmogenic isotope) and (radioactivity, spallation neutrons).
$^8$He/$^9$Li is induced by the $\beta-n$ cascade decay from the interaction between cosmic muon and $^{12}$C in LS, while fast neutron is produced by the coincidence between the energetic neutrion from cosmic muon and the recoiled proton. 
($\alpha$, $n$) backgroud is correlated with the reaction from Thorium and Uranium radioactivities in the detector material and $^{13}$C in LS. 
Based on the information in Ref.~\cite{An:2015jdp}, we estimate the backgroud rate per day\footnote{The reactor neutrinos per day are comprised of 54.71 events from the Yangjiang NPP and Taishan NPP, 2.78 events from the Daya Bay NPP and 1.68 events from the Huizhou NPP.} in table~\ref{tab:bkg}.
\begin{table}[tbp]
 \centering
 \caption{The backgroud rate per day in the nominal JUNO detector.} 
 \label{tab:bkg}
 \begin{tabular}{|c|cccccc|}
  \hline
    Background                    &\makecell[c]{Reactor \\ Neutrino}                 &Accidental                &$^8$He/$^9$Li                &\makecell[c]{Fast \\ Neutron}                &($\alpha$, $n$)              & Geoneutrino \\
    \hline
    Event Rate (day$^{-1}$)       &59.17                            &0.9                       &1.6                          &1.16                          &0.05                         &1.5        \\
    \hline

 \end{tabular}

\end{table}

\section{Monte Carlo event simulation and time-correlation between adjacent events}\label{sec3}
\begin{figure*}[tbp]
\centering
 \includegraphics[width=0.7\textwidth]{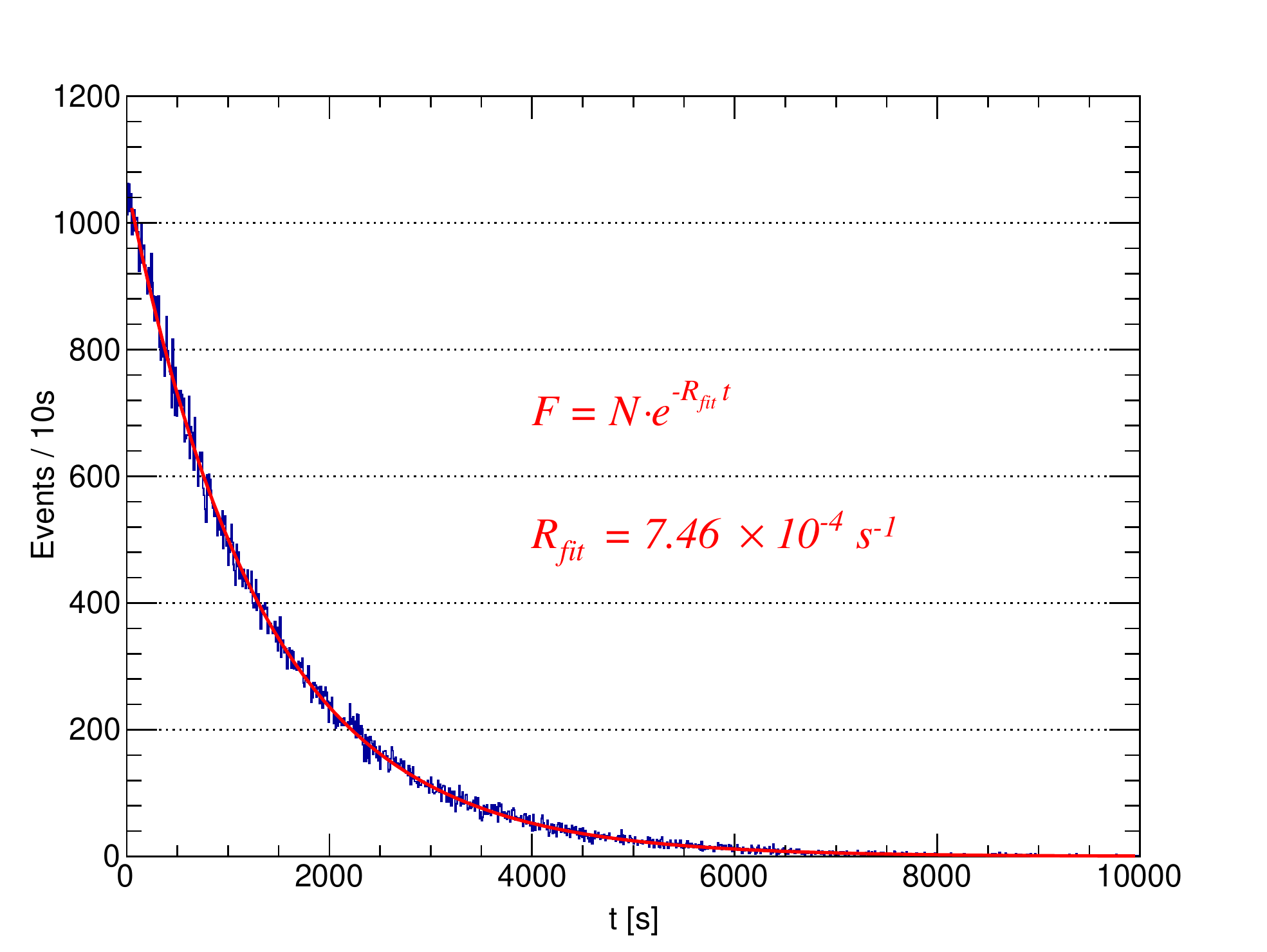} 
\caption{The time intervals between adjacent events, based on a Monte Carlo event simulation for JUNO-like experiment within data-taking of 6 years, the fitted antineutrino event rate ($R_{fit}$) is almost equal to the input antineutrino event rate ($R$).}
\label{fig:time_distribution}
\end{figure*}

For the background estimation in section~\ref{sec2}, neutrino mass ordering issue has no impact on the event rate. In our analysis, we assume the Normal Hierarchy (NH) to be the true mass hierarchy.
Within a $\sim$ 7.45 $\times$ 10$^{-4}$ s$^{-1}$ backgroud rate, we perform a Monte Carlo event simulation corresponding to data-taking of 6 years to produce background status in the nominal JUNO detector. The adjacent backgrouds are expected to follow the Poisson progress within the pdf of $P(0; Rt)$\footnote{In our simulation, $P(0; Rt)$ is randomly produced by the uniform distribution between 0 and 1.}, where $R$ is the background rate. The formula of the pdf is given by:
\begin{align}\label{Eq_PDF}
    P(k;\lambda) = \lambda^k e^{-\lambda} / k!
\end{align}
where $k$ is the observed events and $\lambda$ is the expected events, respectively. After one simulation, we obtain $\sim$ 1.41 $\times$ 10$^5$ backgrouds in total at the JUNO detector and the time intervals between adjacent events follow the exponential distribution, as shown in figure~\ref{fig:time_distribution}. 
Subsequently, we sort all events in the time order to achieve the total number of backgrounds at any period， which can be used to monitor the event excesses at detector.

To identify the potential event excess, we define a time window (cut-off time) to cut off two adjacent backgrounds when the time interval among them is longer than our defined cut-off time. Within such a time window, all events are divided into a variety of bundles, where the event multiplicity is different.
The number of each kind of bundle with the same multiplicity, are measured in the JUNO-like experiment(s) while the expected number is given by:
\begin{align}\label{Eq_Mult}
 N(m) = RTe^{-2R \Delta t}(1 - e^{-R \Delta t})^{m - 1},
\end{align}
where $m$ is the multiplicity, $T$ is the time of data-taking and $\Delta t$ is the cut-off time, respectively. The detailed derivation of eq. \eqref{Eq_Mult} is shown in appendix~\ref{sec:app}. 
The $e^{-2R \Delta t}(1 - e^{-R \Delta t})^{m - 1}$ item represents the probability of $m$ adjacent backgrouds bundled together, where all two adjacent backgrouds are detected inside the cut-off time. Apparently, the probability decreases rapidly by the order of ($m - 1$). 

\begin{figure*}[tbp]
 \centering
 \includegraphics[width=0.45\textwidth]{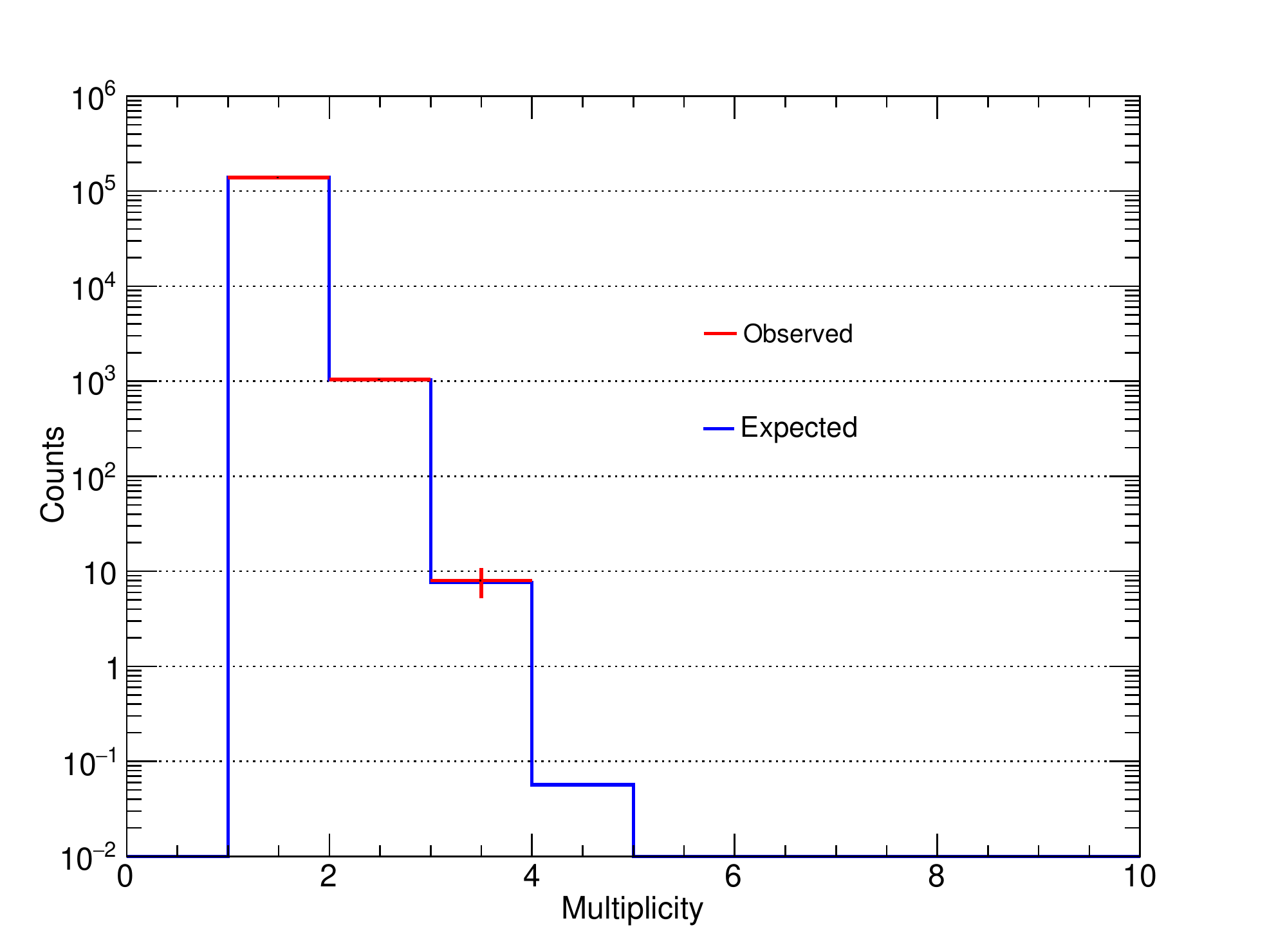}  \quad  \includegraphics[width=0.45\textwidth]{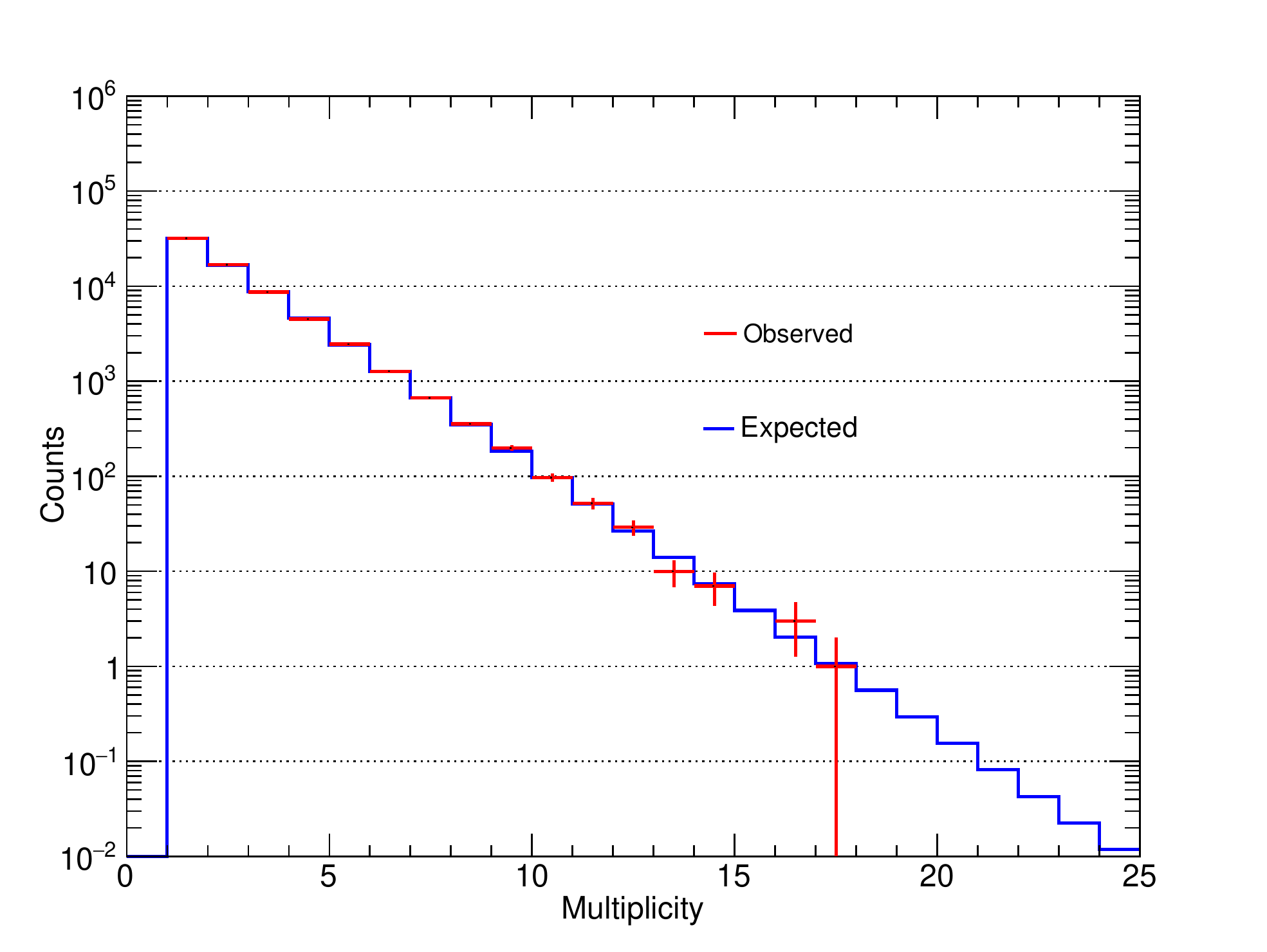} 
\caption{The multiplicity of neutrino events in each kind of bundle divided by requiring the time interval between the adjacent neutrinos smaller than the cut-off time. Left: the cut-off time is given to be 10 s; Right: the cut-off time is given to be 1000 s. The observed values are obtained from the MC simulation and the expected ones are from the theoretical calculation.}
\label{fig:multiplicity}
\end{figure*}

The results of our simulations are shown in figure~\ref{fig:multiplicity}, where the left panel corresponds to the assumption of 10 s' cut-off time ($\Delta t$ = 10). 
This panel shows that for 10s-time-window, $m$ is not expected to be larger than 3, which means that a JUNO-like detector can only observe a bundle of 3 adjacent neutrinos or less, with 6 years of data-taking. 
For a bundle of 4 adjacent backgrouds, the detection probability is found to be 4.03 $\times$ 10$^{-7}$, corresponding to one observation requiring around 106 years data-taking.
Therefore, the medium baseline reactor neutrino experiment(s), has a significant strong capability to identify the event excess within a short time of several tens seconds.
In the future, the neutrinos emitted from gravitational wave sources could give rise to an event excess to the JUNO-like detector, if the corresponding strength is around several events within tens of seconds.

Besides, we also study the impact of the length of cut-off time on the analysis. We change the cut-off time from 10 s to 1000 s and show the results in the right panel of figure~\ref{fig:multiplicity}.
In the absence of astrophysical neutrinos, the observed number is expected to be consistent with the theoretical calculations (the expected values in figure~\ref{fig:multiplicity}) as well, for each multiplicity.
To provide a prudent analysis, we calculate the detection probabilities for different values of $m$. We still convert the probabilities into the period / time interval required for one average observation of a bundle of $m$ adjacent neutrinos like the $m$ = 4 case in 10 s' cut-off time, see the table~\ref{tab:p}. 
For example, the detection probability of a bundle of 24 adjacent neutrinos is found to be 8.35 $\times$ $10^{-8}$, corresponding to 509.82 years for one average observation. Therefore, it is very unlikely to observe such a signal at a short time.
According to our definition, $m$ = 24 implies that the accumulated $\bar\nu_e$ signals emitted from gravitational wave sources can't be identified from backgrouds until over 23 events within the time window of between 1000 s and 24000 s. Otherwise, the signals will be drowned into the backgrounds. 
\begin{table}[tbp]
 \centering
 \caption{The detection probability and required time interval for one average observation within a bundle of $m$ adjacent neutrinos under the 1000 s' cut-off time.} 
 \label{tab:p}
 \begin{tabular}{|c|cccc|}
  \hline
    Multiplicity ($m$)                                                          &21                         &22                        &23                          &24                         \\
    \hline
    Detection Probability                                                       &5.76 $\times$ $10^{-7}$   &3.03 $\times$ $10^{-7}$    &1.59 $\times$ $10^{-7}$     &8.35 $\times$ $10^{-8}$    \\
    \hline
    \makecell[c]{Required Time Interval \\ for One Observation (year)}          &73.88                     &138.53                     &267.79                      &509.82       \\
    \hline
 \end{tabular}

\end{table}

\section{Sensitivity analysis and model constraint on distance of source }\label{sec4}

\subsection{Sensitivity analysis}\label{sec4.1}

As discussed in the previous section, the number of backgrounds is expected to be small and basically constant, with the nominal setup of medium baseline reactor experiment.
With an appropriate cut-off time between two adjacent backgrounds, eq. \eqref{Eq_Mult} can be employed to check whether an event excess over the background level occurs in the daily detector performance, especially in the period when some other messengers have been detected in astrophysical observations.

Without an excess, we can estimate the upper limit on signals by estimating the expected background rate and the observed neutrino candidates, where a classical confidence belt is constructed for this calculation~\cite{Gando:2016zhq,Agostini:2017pfa}.
More generally, one can always perform an analysis on the sensitivity upper limit by estimating the expected background rate, independently. In this section, we make the sensitivity calculation on $\bar\nu_e$ from gravitational wave sources at a JUNO-like experiment, which is evaluated based on the Feldman-Cousins method~\cite{Feldman:1997qc}. 
This method is developed to be an unified approach to estimate the upper limits and sensitivity for small signals and backgrounds, and efficiently resolves the problems resulted from the unphysical confidence belt in the Wilks' theorem~\cite{Wilks:1938ss}.
Due to the low statistical signal candidates and background estimation at the JUNO-like experiment, a maximum likelihood ratio method based on Poisson statistic is adopted
to estimate the upper limit on $\bar\nu_e$ signal:

\begin{align}\label{Eq_chi2}
 \chi^2(\mu) = 2[(\mu + n_\text{bkg}) - n_\text{obs} + n_\text{obs} \cdot \text{ln} \frac{n_\text{obs}}{\mu + n_\text{bkg}}]
\end{align}
where $\mu$ is the expected number of $\bar\nu_e$ signals from potential gravitational wave sources, $n_\text{bkg}$ is the estimated background rate and $n_\text{obs}$ is the total number of observed neutrino candidates.
A test statistics is constructed by:
\begin{equation} 
\begin{split}\label{Eq_dchi2}
 \Delta\chi^2(\mu) & = \chi^2(\mu) - \chi^2(\mu_\text{best}) \\ & = 2[\mu  - \mu_\text{best} + n_\text{obs} \cdot \text{ln} \frac{\mu_\text{best} + n_\text{bkg}}{\mu + n_\text{bkg}}]
\end{split}
\end{equation}
where $\mu_\text{best}$ is the best-fit value. $\Delta\chi^2(\mu)$ represents the deviation between the test neutrino signals $\mu$ and the best-expected neutrino signals $\mu_\text{best}$.

\begin{figure*}[tbp]
\centering
 \includegraphics[width=0.7\textwidth]{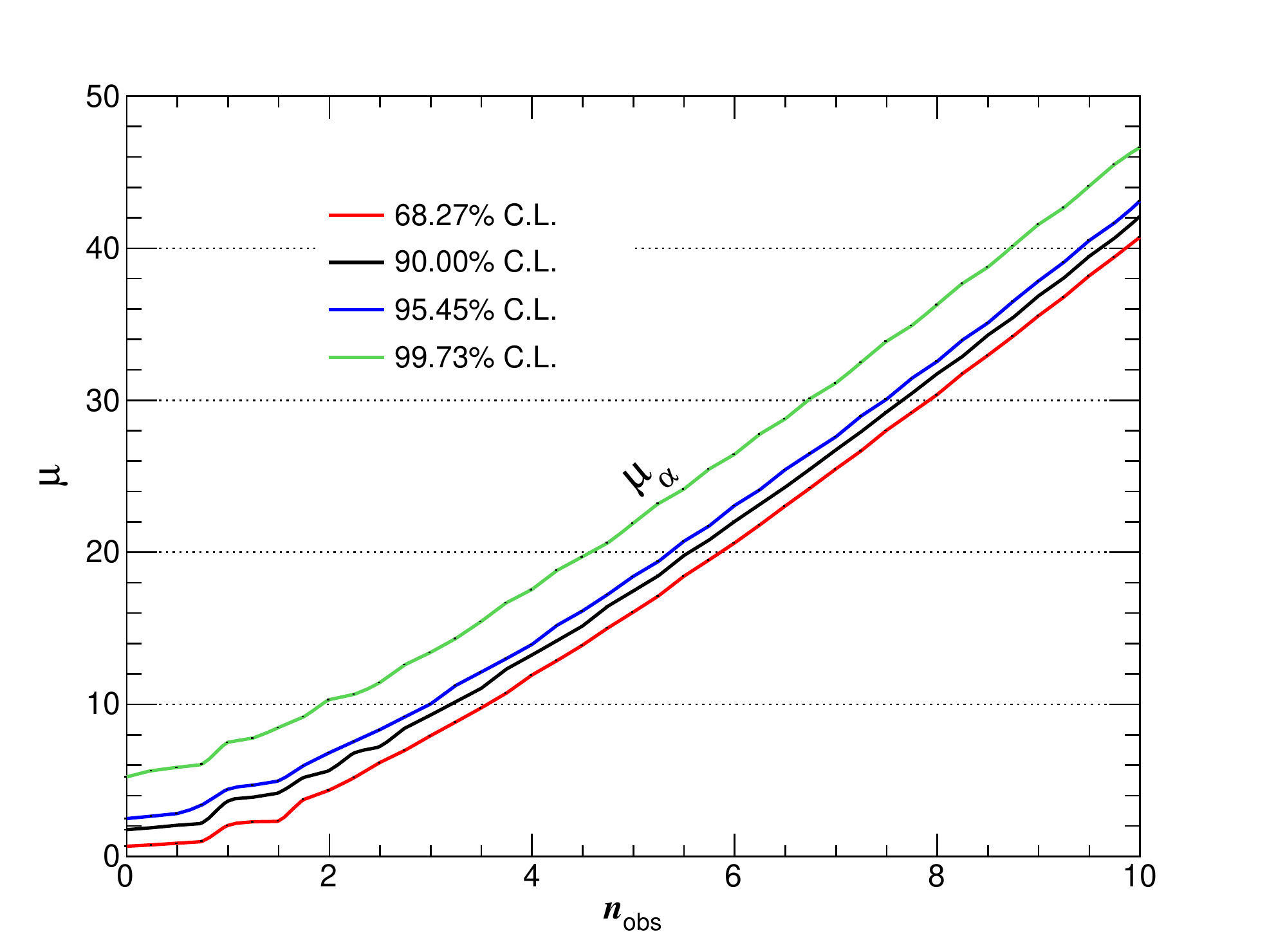} 
\caption{The $\alpha$ percent upper limits on the detected neutrino signals within total measured neutrino candidates inner the time window of 1000 s at a JUNO-like detector. Here, the background rate is based on the above calculation, while the total IBD candidates are supposed to be different value. In real data, the total IBD candidates are expected to be integer.}
\label{fig:FC_sig}
\end{figure*}

On one hand, for each value of $\mu$, a large number of Monte Carlo simulations, are performed to achieve the distribution of $\Delta\chi^2(\mu)$ to specify $\Delta\chi^2_\alpha(\mu)$. $\Delta\chi^2_\alpha(\mu)$ is exactly determined by requiring the ratio of the left integral to the whole integral to be $\alpha$ percent, details can be found in~\cite{Feldman:1997qc}. Actually, 1 - 0.01$\alpha$ represents the significance level.
On the other hand, $\Delta\chi^2_\text{data}(\mu)$, based on the data of real experiment, is compared with $\Delta\chi^2_\alpha(\mu)$. The confidence belt at the acceptance of $\alpha$ percent is constructed as the region where all values of $\mu$ are limitted by $\Delta\chi^2_\text{data}(\mu) \leq \Delta\chi^2_\alpha(\mu)$.
The $\alpha$ percent C.L. upper limit on $\mu$ (which is the expected neutrino signals from gravitational wave sources), is defined as $\mu_\alpha$. In the real experiment, the 90$\%$ C.L. upper limit is generally reported for the ``non-detection" result.
Using the backgroud rate within a time window of 1000 s \footnote{Regarding to $\bar\nu_e$ associated with gravitational wave signals, we adopt a time window of 1000 s, which is originated from the time setup of $\pm$ 500 s for gravitational wave transients in these analyses performed by ANTARRES and IceCube~\cite{Adrian-Martinez:2016xgn,ANTARES:2017iky,ANTARES:2017bia}, KamLAND~\cite{Gando:2016zhq}, Super-Kamiokande~\cite{Abe:2016jwn,Abe:2018mic} and Borexino~\cite{Agostini:2017pfa}.}, we show the upper limits on $\bar\nu_e$ signals at 4 confidence levels with different values of $n_\text{obs}$ in figure~\ref{fig:FC_sig}.
In absence of the real data, we can only estimate the sensitivity upper limit at $\alpha$ percent C.L. when assume $n_\text{obs}$ = $n_\text{bkg}$. At 90$\%$ C.L., the sensitivity upper limit is estimated to be $\mu_{90}$ = 2.44 for a JUNO-like experiment.

\subsection{Model constraint on distance of source}\label{sec4.2}

To evaluate the electron antineutrino flux on the Earth, the upper limit on neutrino signals, $\mu_{\alpha}$, can be translated into the upper limit on fluence, $F^\alpha_\text{UL}$. Without oscillation, this upper limit of neutrinos per cm$^2$ is given by:
\begin{align}\label{Eq_FL}
 F^\alpha_\text{UL} = \frac{\mu_\alpha}{N_\text{T} \int S_\text{n}(E_\nu) \sigma(E_\nu) \epsilon(E_\nu) dE_\nu} = \frac{\mu_\alpha}{N_\text{T} \bar\epsilon \int S_\text{n}(E_\nu) \sigma(E_\nu) dE_\nu}
\end{align}
where $N_\text{T}$ is the total number of target protons, $S_\text{n}(E_\nu)$ is the normalized neutrino energy spectrum, $\sigma(E_\nu)$ is the neutrino cross section, $\epsilon(E_\nu)$ is the total detection efficiency and $\bar\epsilon$ is the total average detection efficiency.
For the nominal JUNO detector, $N_\text{T}$ is calculated to be around 1.45 $\times$ 10$^{33}$ and $\bar\epsilon$ is supposed to be $\sim$ 0.73~\cite{An:2015jdp}. $\sigma(E_\nu)$ is adopted to be the neutrino IBD cross section in ref.~\cite{Strumia:2003zx}, while $S_\text{n}(E_\nu)$ is always adopted to be two different spectra.

In astronomy, the total luminosity can be estimated from the fluence on the Earth, which is given by:
\begin{align}\label{Eq_Lum}
 L = L_s \cdot \Delta T = F \cdot 4 \pi D^{2} \cdot <E>
\end{align}
where $L_s$ is the total luminosity per unit time, $\Delta T$ is the time of neutrino emission, $D$ is the distance between the astrophysical source and the Earth, $<E>$ is the average neutrino energy. With a defined luminosity, the detectable distance can be constrained, conversely.
Regarding to the gravitational wave sources, several models have been employed to simulate the processes of BNS~\cite{Sekiguchi:2015dma,Foucart:2015gaa,Sekiguchi:2016bjd}, BH-NS~\cite{Deaton:2013sla,Foucart:2015vpa} or BBH~\cite{Caballero:2015cpa} mergers and used to evaluate the magnitude of the luminosity and average energy. 
These models are established to emulate the post-merger evolution of gravitational wave sources, which can be described as the radiation-hydrodynamical process in general relativity, using the Einstein's equations and the equations of state (EOS) of hydrodynamics. 
Neutrino leakage schemes~\cite{Sekiguchi:2010ep} are usually used in the simulations of BNS and BH-NS mergers, while the accretion disk models affected by the black hole spin are also adopted into BBH or BH-NS mergers.  
Recently, F. Foucart et al study the neutrino emissions based on the models drived by different EOSs at BNS~\cite{Sekiguchi:2015dma,Foucart:2015gaa,Sekiguchi:2016bjd} and BH-NS~\cite{Deaton:2013sla,Foucart:2015vpa} mergers. O. L. Caballero et al discuss the influence of black hole spin on accretion disk neutrino detection and use a set of accretion tori to produce neutrinos~\cite{Caballero:2015cpa}.
The corresponding luminosities and average energy for $\bar\nu_e$ are shown in the table~\ref{tab:LE} in these models, details can be referred to refs.~\cite{Sekiguchi:2015dma,Foucart:2015gaa,Sekiguchi:2016bjd,Deaton:2013sla,Foucart:2015vpa,Caballero:2015cpa}. For the accretion tori, the calculations based on the case of one single black hole with a mass of 3 $M_\odot$~\cite{Caballero:2015cpa} are performed.

\begin{table}[tbp]
 \centering
 \caption{The luminosity and average energy for $\bar\nu_e$ based on different models at BNS or BBH or BH-NS mergers.} 
 \label{tab:LE}
 \begin{tabular}{|c|cccc|cccc|}
  \hline
    \multirow{2}{*}{Model}                   &\multicolumn{4}{c|}{EOSs}                       &\multicolumn{4}{c|}{Accretion Tori}  \\
                                             &SFHo       &LS220      &DD2     &LS            &J0       &C0      &Ja      &Ca      \\
                       
    \hline  
    Source                                   &\multicolumn{3}{c}{BNS}         &BH-NS         &\multicolumn{4}{c|}{BBH/BH-NS}     \\
    Mass ($M_\odot$)                         &\multicolumn{3}{c}{1.2 + 1.2 }  &5.6 + 1.4     &\multicolumn{4}{c|}{3}             \\
    \hline
    $L_s$ ($\times$ 10$^{53}$ erg s$^{-1}$)  &3.0        &2.1        &2.2     &5.0           &2.7      &4.8     &3.7     &9.8    \\
    $<E>$ (MeV)                              &14.5       &13.8       &13.8    &15            &12.7     &10.3    &13.4    &11.8     \\
    \hline

 \end{tabular}

\end{table}

According to the above models, the magnitude of luminosities are estimated to be around 10$^{53}$ erg s$^{-1}$ after the merger time of $\sim$ 10 ms while the average neutrino energy ranges from 10 to 15 MeV within small differences.
These differences are considered to be dependent on the parameters of model and the masses of source. Typically, considering the neutrino cooling process at a hypermassive neutron star, the cooling time is estimated to be around 2 - 3 s~\cite{Sekiguchi:2011zd,Kiuchi:2012mk}. Therefore, we set up the $\Delta T$ to be 3 s in eq. \eqref{Eq_Lum} for BNS and BH-NS mergers. However, for BBH mergers, there are no estimations on the emission time of neutrinos yet. For consistence, we assume the emission time of BBH mergers is also 3 s. 
To constrain the detectable distance, we have to first estimate the fluence on the Earth. Up to now, in the absence of known mechanism for the emission of neutrinos in BBH, BNS and BH-NS mergers, we can empirically assume neutrinos are monochromatic or follow the Fermi-Dirac distributions~\cite{Gando:2016zhq,Abe:2016jwn,Abe:2018mic,Agostini:2017pfa}.

In the assumption of monochromatic energy spectrum (McES), neutrinos are expected to be mono-energetic with the same energy $E_\nu$ and eq. \eqref{Eq_FL} is simplified as:
\begin{align}\label{Eq_FLm}
 F^\alpha_\text{UL} = \frac{\mu_\alpha}{N_\text{T} \bar\epsilon \sigma(E_\nu)}
\end{align}
eq. \eqref{Eq_FLm} provides the most conservative (or largest) upper limit on the electron antineutrino fluence. As the IBD cross section is proportional to $E_\nu^2$, the fluence is inversely proportional to $E_\nu^2$.
\begin{figure*}[!htbp]
\centering
 \includegraphics[width=0.7\textwidth]{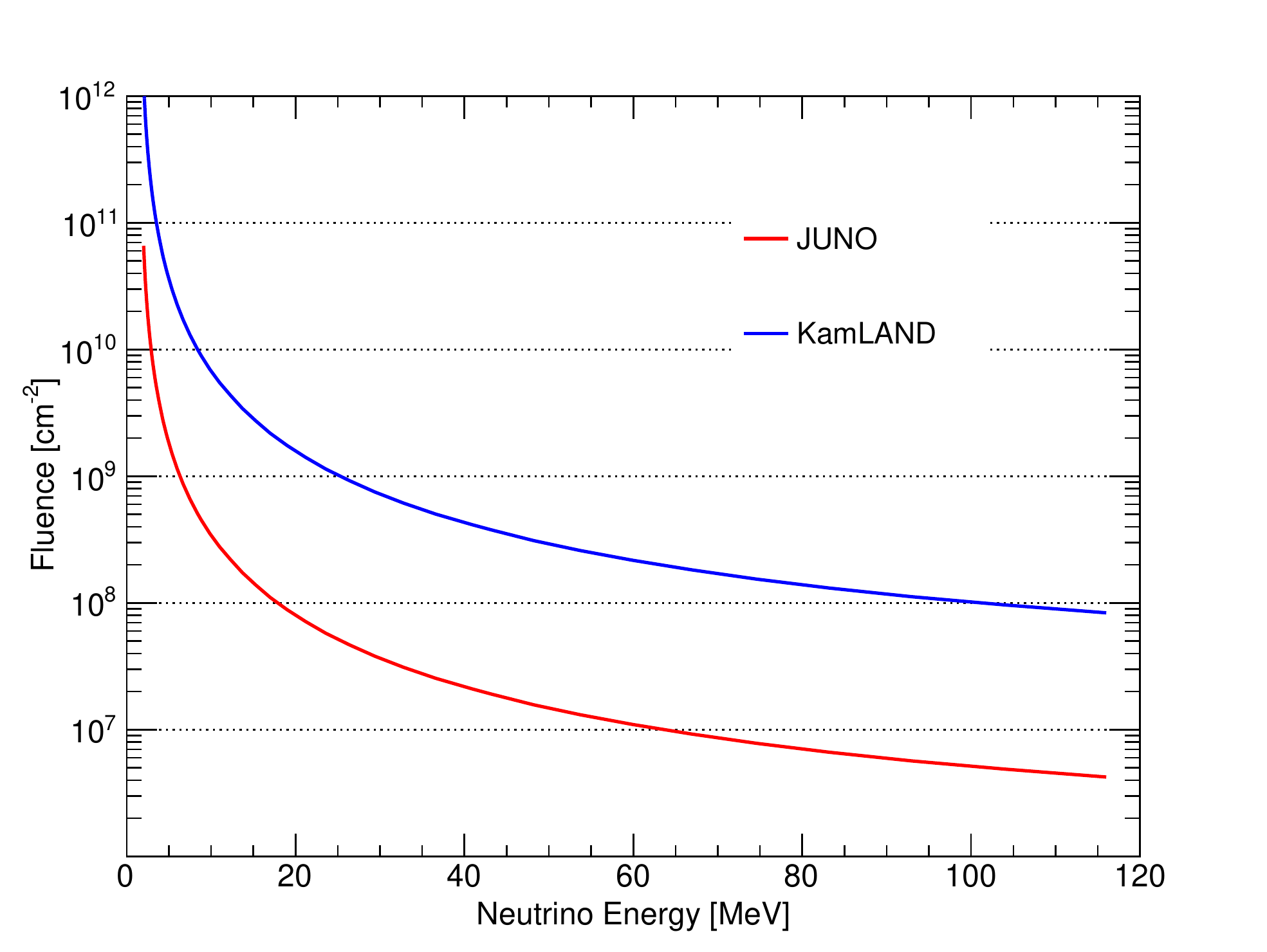} 
\caption{The red line is the sensitivity on the electron antineutrino fluence at 90$\%$ C.L., which represents the most conservative detection capability for neutrino flux within the time window of 1000 s on the Earth at a JUNO-like detector. As a comparison, the corresponding sensitivity of the KamLAND detector is also shown and represented by the blue curve. The calculations are based on the event rate in ref.~\cite{Gando:2016zhq}.}
\label{fig:FL_Onedet}
\end{figure*}

With respect to sensitivity at 90$\%$ C.L. ($\mu_{90}$), the electron antineutrino fluence ($F_\text{UL}^{90}$) is constrained in the range from about $6 \times 10^{10}$ cm$^{-2}$ to about $4 \times 10^{6}$ cm$^{-2}$ for the neutrino energy range from 1.8 MeV to 120 MeV, corresponding to the red curve in figure~\ref{fig:FL_Onedet}. 
It implies $\bar\nu_e$ with such that flux on the Earth can be distinguished at 90$\%$ C.L. from  $\bar\nu_e$ backgrounds at the JUNO-like detector, where the fluence sensitivity is around twenty times better than KamLAND\footnote{For KamLAND, we only use the efficiency and event rate at the GW151226 period in ref.~\cite{Gando:2016zhq} and ignore the differences between each period. In fact, the differences can make some small impacts on the sensitivity.} because of the significant increment of target mass.   
Table~\ref{tab:FD} shows the estimated fluence ($F_\text{UL}^{90}$) and constraints of the detectable distance ($D_\text{UL}^{90}$) based on different models in table~\ref{tab:LE}. Basically, the JUNO-like detector can observe the monochromatic neutrinos from gravitational wave sources with the distance of 1 - 3 Mpc at 90$\%$ C.L., which is about four times further than KamLAND.

On the other hand, neutrinos are also usually supposed to obey the normalized Fermi-Dirac distribution for zero chemical potential $\eta = 0$, which is given by:
\begin{align}\label{Eq_FD}
 S_\text{n}(E_\nu) = \text{F}_\text{norm} \cdot \frac{E_\nu^2}{1 + e^{E_\nu / T - \eta}}
\end{align}
where F$_\text{norm}$ is the normalization factor and $T$ is the effective neutrino temperature, which is given by $T \approx <E> / 3.1514$~\cite{Keil:2002in}. The normalized Fermi-Dirac distributions are slightly different among the models in table~\ref{tab:LE} due to the differences of $<E>$.

With respect to sensitivity at 90$\%$ C.L. ($\mu_{90}$), the electron antineutrino fluences ($F_\text{UL}^{90}$) are calculated by substituting the Fermi-Dirac energy spectrum (FDES) into eq. \eqref{Eq_FL} and performing integration within the energy range from 1.8 to 120 MeV, as shown in table~\ref{tab:FD}.  
Furthermore, the detectable distances at 90$\%$ C.L. ($D_\text{UL}^{90}$) are also shown in table~\ref{tab:FD}. For JUNO-like detector, $F_\text{UL}^{90}$ are generally within 1 - 3 $\times$ 10$^8$ cm$^{-2}$ and the corresponding $D_\text{UL}^{90}$ are calculated to be around 1 - 3 Mpc.
However, for the KamLAND detector, $F_\text{UL}^{90}$ is estimated to be around twenty times weaker than JUNO-like detector while $D_\text{UL}^{90}$ is obtained to be about a quarter distance of the one of JUNO-like detector.

\begin{table}[tbp]
 \centering
 \caption{The fluence ($\times$ 10$^8$ cm$^{-2}$) and detectable distance (Mpc) at 90$\%$ C.L. for $\bar\nu_e$ with the average energy based on different models at BNS or BBH or BH-NS mergers, at the JUNO and KamLAND experiment.} 
 \label{tab:FD}
 \begin{tabular}{|ccc|cccc|cccc|}
  \hline
    \multicolumn{3}{|c|}{\multirow{2}{*}{Model}}                   &\multicolumn{4}{c|}{EOSs}                       &\multicolumn{4}{c|}{Accretion Tori}  \\
    \multicolumn{3}{|c|}{}            &SFHo       &LS220      &DD2     &LS            &J0       &C0      &Ja      &Ca  \\
    \hline
    \multirow{4}{*}{JUNO}       &\multirow{2}{*}{$F_\text{UL}^{90}$}      &McES    &1.56       &1.72       &1.72    &1.45     &2.07     &3.24    &1.84    &2.42       \\
                                &                                         &FDES    &1.02       &1.13       &1.13    &0.94     &1.35     &2.12    &1.20    &1.58        \\
                                &\multirow{2}{*}{$D_\text{UL}^{90}$}      &McES    &1.44       &1.18       &1.21    &1.90     &1.27     &1.50    &1.53    &2.32      \\
                                &                                         &FDES    &1.79       &1.45       &1.49    &2.35     &1.57     &1.86    &1.90    &2.87       \\
                                \hline
    \multirow{4}{*}{KamLAND}    &\multirow{2}{*}{$F_\text{UL}^{90}$}     &McES    &30.84      &33.89      &33.89   &28.65   &40.86    &69.97   &36.29   &47.87        \\
                                &                                        &FDES    &20.06      &22.29      &22.29   &18.66   &26.63    &41.87   &23.74   &31.20     \\
                                &\multirow{2}{*}{$D_\text{UL}^{90}$}     &McES    &0.32       &0.27       &0.27    &0.43     &0.29     &0.34    &0.35    &0.52        \\
                                &                                        &FDES    &0.40       &0.33       &0.33    &0.53     &0.35     &0.42    &0.43    &0.65       \\
    \hline

 \end{tabular}

\end{table}

\section{An extra detector}\label{sec5}

As discussed in section~\ref{sec4}, due to the large target mass and low background rate, a JUNO-like detector is expected to provide a better platform to the search of neutrinos from gravitational wave sources than any other existing reactor neutrino experiments.
In order to efficiently improve the sensitivity, besides simply increasing the scale of the detector, building an extra detector could be a more realistic option. It could be more effective to improve the sensitivities and constraints to the small signals after the original JUNO-like detector starts data-taking.

Regarding to the extra detector, since the expected neutrino signals ($\mu$) could be different with the original JUNO-like detector due to the discrepancies on target mass and detection efficiency between two detectors. 
The estimation of the sensitivity is more complicated than just repeating the analysis in section~\ref{sec4}. Instead, we establish a multi-detectors' analysis that is constructed by the combination of two (multiple) detectors and the Feldman-Cousins Method is applied again. 
To attain the fluence from two detectors with different target mass and detection efficiency, we define a parameter to represent the expected anti-neutrino signals per effective proton, which is given by:
\begin{align}\label{Eq_muC}
 \mu_\text{C} = F \cdot \int S_\text{n}(E_\nu) \sigma(E_\nu) dE_\nu 
\end{align}
Therefore, from eq. \eqref{Eq_chi2} and eq. \eqref{Eq_FL}, we have 
\begin{align}\label{Eq_chi2new}
 \chi^2(\mu_\text{C}) = \sum_i^{dets} 2[(\mu_\text{C} \cdot N_{\text{T}, i} \cdot \bar\epsilon_i + n_{\text{bkg}, i}) - n_{\text{obs}, i} + n_{\text{obs}, i} \cdot \text{ln} \frac{n_{\text{obs}, i}}{\mu_\text{C} \cdot N_{\text{T}, i} \cdot \bar\epsilon_i + n_{\text{bkg}, i}}]
\end{align}
Actually, $\mu_\text{C}$ is same for all detectors, which represents the expected neutrino signals per effective proton. 
For one single JUNO detector, $\mu_\text{C}$ = 2.31 $\times$ 10$^{-33}$ at 90$\%$ C.L., corresponding to $\mu_{90}$ = 2.44, which is consistent with the result of single detector. Furthermore, using two assumed energy spectra, we can obtain the same results with section~\ref{sec4}, respectively.

\begin{figure*}[!htbp]
\centering
 \includegraphics[width=0.45\textwidth]{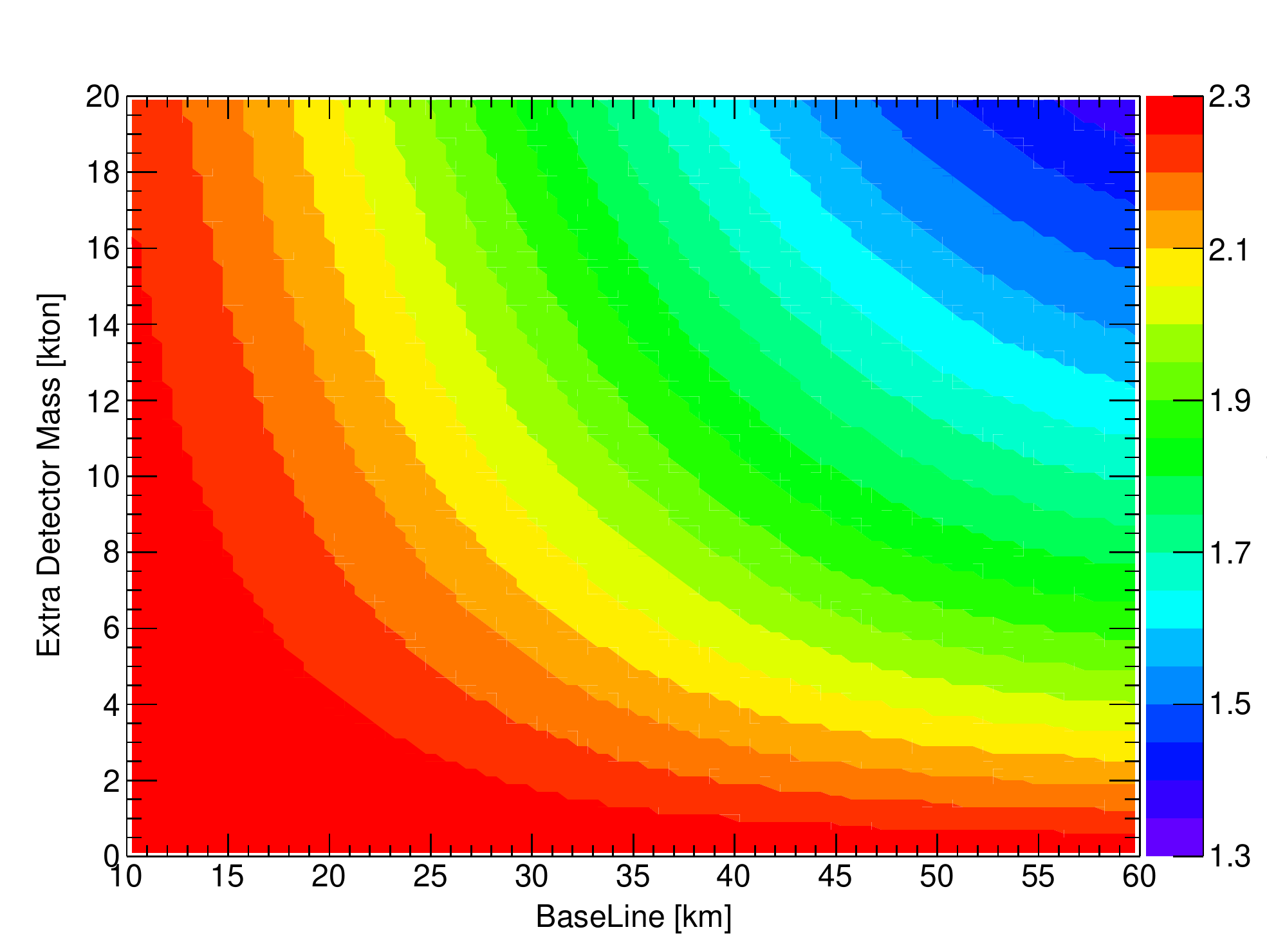}  \quad   \includegraphics[width=0.45\textwidth]{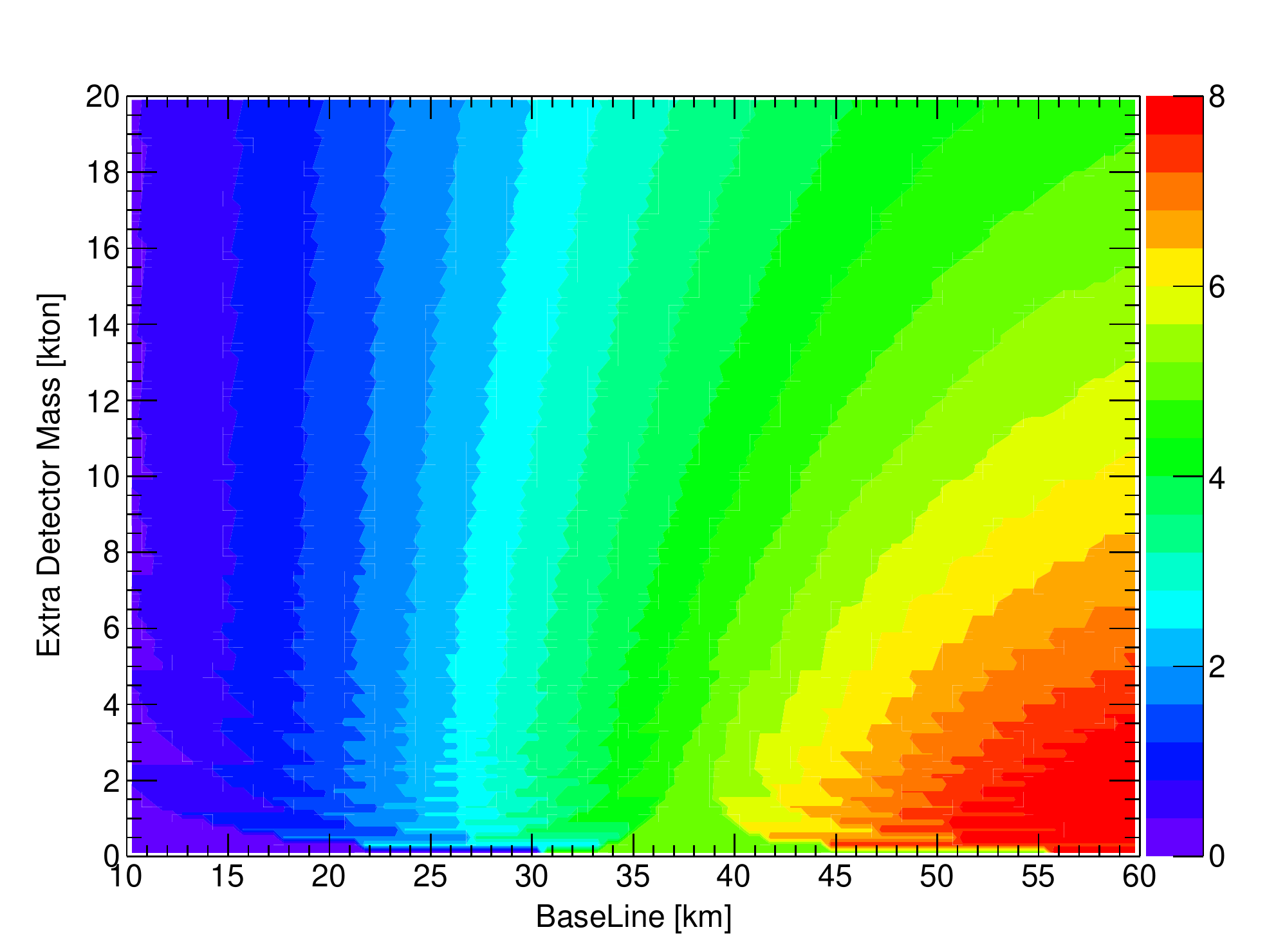}
\caption{Left: The upper limit on $\bar{\nu}_e$ at 90$\%$ C.L. ($\mu_\text{C}^{90}$)as a function of target mass and baseline of the extra detector. The legend of color code is shown on the right bar, which represents $\mu_\text{C}^{90}$ per effective proton $\times$ 10$^{-33}$. Right: The decrement of $\mu_\text{C}^{90}$ at per unit target mass ($R_{\mu_\text{C}}^{90}$) as a function of target mass and baseline of the extra detector. The legend of color code is shown on the right bar, which represents $R_{\mu_\text{C}}^{90}$ $\times$ 10$^{-35}$ kton$^{-1}$.}
\label{fig:Sig_Twodet}
\end{figure*}

Regarding to the systematics of the extra detector, we assume that it is identical to the nominal JUNO-like detector except the target mass and distance to the reactor cores. 
On one hand, we ignore the environmental impacts among two detectors and assume all backgrouds discussed in section~\ref{sec2} are proportional to the target mass of the extra detector.
On the other hand, the reactor neutrino background is assumed to be dependent on the distance between the detector and reactor core as well as discussed in section~\ref{sec2}. 
Moreover, if an extra detector is proposed in the JUNO experiment, the optimal location should be at a distance $>$ 200 km away from the Daya Bay and Huizhou NPP. In order to do a conservative but prudent simulation, we assume this distance to be 200 km to obtain a maximum value of reactor background rate from the Daya Bay and Huizhou NPP, which is calculated to be 0.325 per kton per day.
Varing the baseline away from Yangjiang and Taishan NPP and target mass, the corresponding $\mu_\text{C}$ at 90$\%$ C.L. ($\mu_\text{C}^{90}$) is calculated and shown in the left panel of figure~\ref{fig:Sig_Twodet}.
Basically, if the extra detector is located in the distance of $>$ 10 km away from Yangjiang and Taishan NPP, $\mu_\text{C}^{90}$ decreases as the target mass and baseline increases.
Since $\mu_\text{C}^{90}$ is proportional to $F_\text{90}$, this implies that with appropriate target mass and baseline, the extra detector provides significant improvement on the constraints to the neutrino fluence on the Earth, which results in better sensitivity.

Regarding to the impact of baseline, a further distance from the reactor core implies a smaller background rate, makes the extra detector more sensitive to neutrinos from gravitational wave sources and thus produces a better constraint on the fluence. 
On the other hand, a larger target mass also leads to a better sensitivity because it is inversely proportional to $\mu_\text{C}$. However, the decrement of $\mu_\text{C}$ becomes slower as the target mass becomes larger.
To quantify the relationship between the decrement of $\mu_\text{C}$ and the target mass, we define a parameter $R_{\mu_\text{C}}$, which is given by:

\begin{align}\label{Eq_R}
 R_{\mu_\text{C}} = \frac{|\Delta\mu_\text{C}|}{M_\text{ext}}
\end{align}
where $M_\text{ext}$ is the target mass of extra detector. The resulting $R_{\mu_\text{C}}$ at 90$\%$ C.L. is shown in the right panel of figure~\ref{fig:Sig_Twodet}. 
In summary, $R_{\mu_\text{C}}^{90}$ decreasing as the target mass increases, indicates the benefit rate (the additional constraints on the neutrino fluence) from the extra detector becomes weaker and weaker.
Combining the left and right panels of figure~\ref{fig:Sig_Twodet} revelas that larger target mass of the extra detector of course gives rises to better constraint (smaller $\mu_\text{C}$).
Nevertheless, the benefits from the extra detector (the additional constraints on the neutrino fluence) also becomes weaker as the right panel shows that the decrement of $\mu_\text{C}$ is small when the targe mass is large.

\begin{figure*}[tbp]
\centering
 \includegraphics[width=0.7\textwidth]{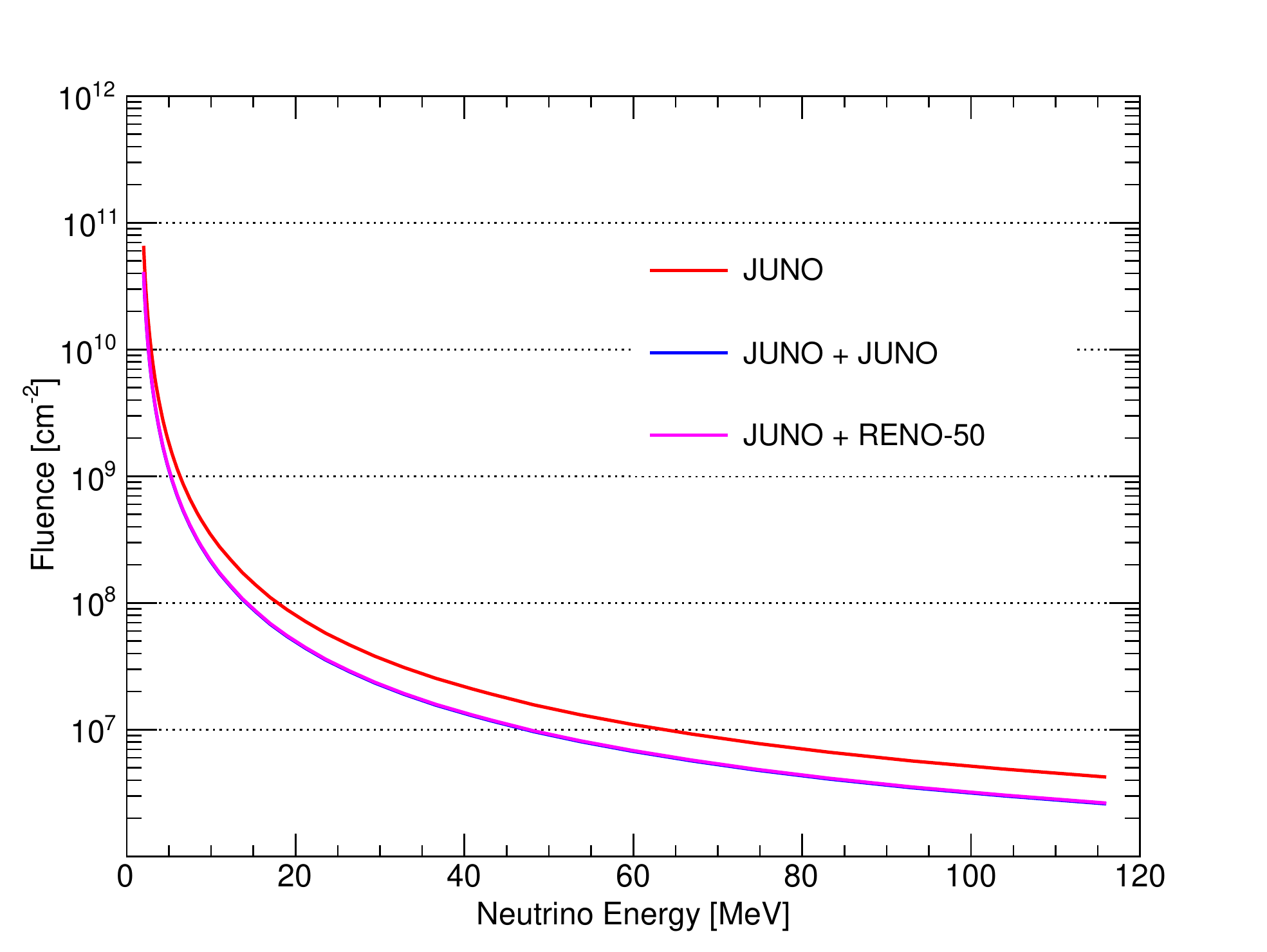} 
\caption{The red line is the upper limits on $\bar{\nu}_e$ fluence at 90$\%$ C.L. at one single JUNO detector, the blue one is the upper limits at two identical JUNO detector while the magenta one is the upper limits at one nominal JUNO detector and one nominal RENO-50 detector. The fluence has a $\sim$ 38$\%$ decrement, corresponding to a $\sim$ 38$\%$ improvement of the sensitivity, when add one identical JUNO detector or combined with the RENO-50 detector in the future.}
\label{fig:FL_Twodet}
\end{figure*}

Particularly, in our simulation, we assume the extra detector to be identical to the original JUNO-like detector. Namely, the extra detector has the same systematics, resolution, target mass and same baseline (but of course locating at a different place) with the original JUNO-like detector.
With such an extra detector, $\mu_\text{C}^{90}$ is changed from 2.31 $\times$ 10$^{-33}$ to 1.42 $\times$ 10$^{-33}$, which implies that the additional JUNO-like detector provides $\sim$ 38.53$\%$ extra constraints to the neutrino fluence. 
With the assumption of monochromatic energy spectrum, the resulting $F_\text{UL}^{90}$ is inversely proportional to neutrino energy, which is represented by the blue curve in the figure~\ref{fig:FL_Twodet}. 
Baesd on the models in table~\ref{tab:LE}, we can also estimate the corresponding fluence and constrain the detectable distance at 90$\%$ C.L. with both the monochromatic assumption and the Fermi-Dirac assumption, as shown in table~\ref{tab:FDE}.
We find that with an additional JUNO-like detector, the detectable distance can be improved by 27.54$\%$.

On the other hand, instead of building an extra detector, if RENO-50 is built in the future, a combined analysis of the data from JUNO and RENO-50 could be an alternative to improve the sensitivity. RENO-50 is the other proposal of medium baseline reactor neutrino experiment, which plans to build a $\sim$ 18 ktons detector, located in Mt. GuemSeong within a beseline of $\sim$ 47 km from the Hanbit NPP of Yonggwang with total thermal power of 16.8 GW$_\text{th}$~\cite{Kim:2014rfa,Seo:2016uom}.
After RENO-50 is built, it could also contribute in the search for neutrinos from gravitational wave sources and make contributions to constraining neutrino fluence in the future. 
The detector of RENO-50 can be equivalently treated as an extra detector of the JUNO detector, and provides significant extra constraints to the neutrino fluence.
   
Regarding to the RENO-50 experiment, in the absence of available background data, we assume that besides the reactor neutrino background, all other backgrounds are proportional to the RENO-50 target mass and have the same magnitude per unit target mass with JUNO. The reactor neutrinos are estimated by the RENO-50 experimental setups. 
Similar to our previous discussion about JUNO, we believe the reactor neutrinos are the major background and have a significant impact on determining the sensitivity. For the RENO-50 experiment, the average detection efficiency is assumed to be $\sim$ 72.6$\%$ according to the RENO collaboration\footnote{In reality, the detection efficiency could be different with the assumption of the RENO collaboration. In absence of the true data, we simply suppose the RENO-50 detector has the same efficiency with the RENO detector.}~\cite{Seo:2016uom}.
Since the thermal power is just about half small of the JUNO experiment, the reactor neutrino background is estimated to be around 50$\%$ of the one of JUNO. 
The corresponding $\mu_\text{C}^{90}$ is calculated to be 1.44 $\times$ 10$^{-33}$, a few larger than one additional JUNO-like detector and $\sim$ 37.66$\%$ smaller than one single JUNO-like detector. 
Finally, $\mu_\text{C}^{90}$ can be converted into not only the resulting $F_\text{UL}^{90}$ with respect to monochromatic energy spectrum, which is represented as the magenta line in figure~\ref{fig:FL_Twodet}, but also $F_\text{UL}^{90}$ and $D_\text{UL}^{90}$ with both the monochromatic assumption and the Fermi-Dirac assumption for a variety of models, which is listed in table~\ref{tab:FDE}.
Generally, an RENO-50 detector can provide a bit worse but almost consistent constraint to neutrino fluence than an extra JUNO detector even if target mass is 2 ktons smaller. In summary, we can obtain an unprecedented sensitivity on the astrophysical neutrino by combining JUNO and RENO-50 experiment in the future.

\begin{table}[tbp]
 \centering
 \caption{The fluence ($\times$ 10$^8$ cm$^{-2}$) and detectable distance (Mpc) at 90$\%$ C.L. for $\bar\nu_e$ with the average energy based on different models at BNS or BBH or BH-NS mergers, at the JUNO + JUNO and JUNO + RENO-50 setups.} 
 \label{tab:FDE}
 \begin{tabular}{|ccc|cccc|cccc|}
  \hline
    \multicolumn{3}{|c|}{\multirow{2}{*}{Model}}                   &\multicolumn{4}{c|}{EOSs}                       &\multicolumn{4}{c|}{Accretion Tori}   \\
    \multicolumn{3}{|c|}{}            &SFHo       &LS220      &DD2     &LS            &J0       &C0      &Ja      &Ca  \\
    \hline
    \multirow{4}{*}{JUNO+JUNO}   &\multirow{2}{*}{$F_\text{UL}^{90}$}      &McES    &0.96       &1.05       &1.05    &0.89     &1.27     &1.99    &1.13    &1.49       \\
                                   &                                         &FDES    &0.62       &0.69       &0.69    &0.58     &0.83     &1.30    &0.74    &0.97     \\
                                   &\multirow{2}{*}{$D_\text{UL}^{90}$}      &McES    &1.84       &1.50       &1.54    &2.42     &1.62     &1.92    &1.96    &2.96      \\
                                   &                                         &FDES    &2.28       &1.85       &1.90    &3.00     &2.00     &2.37    &2.42    &3.66      \\
    \hline
    \multirow{4}{*}{JUNO+RENO-50}&\multirow{2}{*}{$F_\text{UL}^{90}$}      &McES    &0.97       &1.07       &1.07    &0.90     &1.29     &2.02    &1.14    &1.51         \\
                                   &                                         &FDES    &0.63       &0.70       &0.70    &0.59     &0.84     &1.32    &0.75    &0.98        \\
                                   &\multirow{2}{*}{$D_\text{UL}^{90}$}      &McES    &1.83       &1.49       &1.53    &2.40     &1.61     &1.90    &1.94    &2.94        \\
                                   &                                         &FDES    &2.26       &1.84       &1.88    &2.98     &1.99     &2.35    &2.40    &3.64       \\
    \hline

 \end{tabular}

\end{table}

\section{Conclusion}\label{sec6}
In this paper, we introduce a method to search for the neutrino excess at the nominal JUNO experiment, which is with the largest target mass among all existing reactor neutrino experiments. This method can be used to compare the observed neutrinos with the expected backgrounds at any period by classifying these neutrinos within a given cut-off time.
With an appropriate cut-off time, we can easily discriminate if there is an event excess. In astronomy, an excess could be associated with astrophysical phenomena. Therefore, we calculate the sensitivity on $\bar{\nu}_e$ signals to evaluate the detection capability of neutrinos from potential gravitational wave sources at one JUNO-like experiment.
Within the nominal setups, we achieve the upper limit of sensitivity on $\bar{\nu}_e$ signal at 90$\%$ C.L. ($\mu_{90}$) of 2.44, which corresponds to fluence range from about 6 $\times$ 10$^{10}$ cm$^{-2}$ to about 4 $\times$ 10$^{10}$ cm$^{-2}$ for the neutrino energy range from 1.8 MeV to 120 MeV at monochromatic energy spectrum assumption.
Using the predicted luminosities in known models describing BNS, BH-NS or BBH mergers, the fluences are constrained to be around 1 - 3 $\times$ 10$^{8}$ cm$^{-2}$ for both monochromatic energy spectrum assumption and Fermi-Dirac energy spectrum assumption, which corresponds to the gravitational wave sources are able to be detected at the distance of about 1 - 3 Mpc. 
Compared with the KamLAND experiment, the sensitivity of the (future) JUNO-like experiment is expected to improve by a factor of twenty.

We further discuss the methods to improve the sensitivity of JUNO-like experiment in the study of neutrinos from gravitational wave sources. The most straightforward way is to design a larger detector to improve the sensitivity. 
On the other hand, we can also construct one or more extra detectors to improve the sensitivity. The sensitivity will be improved as the detected $\bar{\nu}_e$ signal per effective proton at 90$\%$ C.L., $\mu_\text{C}^{90}$, decreases by baseline and target mass as shown in figure~\ref{fig:Sig_Twodet}.
Particularly, when the extra detector is designed to be identical to the original JUNO detector, the sensitivity will become $\sim$ 38.53$\%$ better than just one single detector and the detectable distance will have $\sim$ 27.54$\%$ improvement correspondingly. 
We also consider another proposed reactor neutrino experiment, RENO-50 and combine it with the JUNO experiment. In this case, we find that the sensitivity will be improved around $\sim$ 37.66$\%$, which is very similar to the case of building an extra detector identical to JUNO.

To make an extensive research, we can develop the analysis method to evaluate the detection capability of neutrinos from the broad astrophysical sources at the future reactor neutrino experiments. These sources are inclusive of the stars, supernovae, gamma-burst sources, black holes, etc. 
However, the appropriate models are crucial since the analyses are always model-dependent. As discussed in this paper, the detectors at the future reactor neutrino experiments can be employed to calculate the neutrino fluence and constrain the detection distance of the astrophysical sources along with the predictions of models.

\appendix
\section{Derivation of the multiplicity probability of a bundle of events}\label{sec:app}
As described in section~\ref{sec3}, the multiplicity ($m$) is counted to be the number of each bundle of neutrinos, which depends on the cut-off time ($\Delta t$) and the time interval between adjacent neutrinos.
If $m$ is expected to be 1, one neutrino is independently divided into a bundle when there is no any one neutrino inner $\Delta t$ before and after it.
According to the Poisson progress, the probability is given by:
\begin{align}\label{Eq_prob1}
P(m = 1) = P(0; R \Delta t)P(0; R \Delta t) = e^{-2R \Delta t}
\end{align}
If $m$ is expected to be 2, two adjacent neutrinos are independently divided into a bundle when there is no any one neutrino inner $\Delta t$ before the first one and after the second one.
Additionally, these two neutrinos are required to have a time interval of < $\Delta t$. Therefore, the probability can be expressed as:
\begin{align}\label{Eq_prob2}
P(m = 2) = P(0; R \Delta t)P(0; R \Delta t) \int_{0}^{\Delta t} P(0; R t_1) P(1; R dt_1)dt_1 = e^{-2R \Delta t}(1 - e^{-R \Delta t})
\end{align}
where $t_1 + dt_1$ is the time interval between these two neutrinos, the integral represents the total probability of these two neutrinos.
Similarly, if $m$ is expected to be 3, the probability is given by:
\begin{align}
\begin{split}\label{Eq_prob3}
P(m = 3) & = P(0; R \Delta t)P(0; R \Delta t) \int_{0}^{\Delta t} P(0; R t_1) P(1; R dt_1)dt_1 \int_{0}^{\Delta t} P(0; R t_2) P(1; R dt_2)dt_2 \\ & = e^{-2R \Delta t}(1 - e^{-R \Delta t})^2
\end{split}
\end{align}
In general, the probability can be expressed as:
\begin{align}
\begin{split}\label{Eq_prob}
P(m) & = P(0; R \Delta t)P(0; R \Delta t) \int_{0}^{\Delta t} P(0; R t_1) P(1; R dt_1)dt_1 \cdots \int_{0}^{\Delta t} P(0; R t_{m - 1}) P(1; R dt_{m - 1})dt_{m - 1} \\ & = e^{-2R \Delta t}(1 - e^{-R \Delta t})^{m - 1}
\end{split}
\end{align}
Finally, the expected number of multiplicity can be referred to eq. \eqref{Eq_Mult}.

\acknowledgments

The authors thank to Jiajie Ling and JiaJun Liao for informative discussions and suggestions.


\end{document}